\documentclass[12pt]{article}
\usepackage{amsfonts}
\usepackage[dvips]{graphicx}
\usepackage{latexsym,amsmath,amssymb,graphics,stmaryrd}
\usepackage{array}
\usepackage{epsfig}
\usepackage{subfigure}

\usepackage{multirow}
\usepackage[dvips]{graphicx}
\usepackage[dvips]{color}
\usepackage{pst-node}
\usepackage{dsfont}
\usepackage{amsthm}
\usepackage{graphics}
\usepackage{subfigure}
\usepackage{fancyhdr}
\usepackage[dvips]{color}

\newcmykcolor{lightorange}{0 0.1 0.2 0}
\newcmykcolor{redorange}{0 0.75 1 0}
\newcmykcolor{lightred}{0 0.125 0.125 0}
\newcmykcolor{lightgreen}{0.125 0 0.125 0}
\newcmykcolor{lightcyan}{0.25 0 0 0}
\newcmykcolor{lightmagenta}{0 0.25 0 0}
\newcmykcolor{lightredorange}{0 0.125 0.133 0}
\newcmykcolor{cyan}{1 0 0 0.2}
\newgray{hiddengray}{0.9}

\newcommand{\col}{~,}
\newcommand{\pnt}{~.}
\newcommand{\AdS}{\text{AdS}}

\newcommand{\twob}{{\text{II}\,\text{B}}}

\newcommand{\N}{\mathcal{N}}

\newcommand{\e}{\operatorname{e}}

\newlength{\neglength}

\DeclareMathOperator{\tr}{tr}

\numberwithin{equation}{section}
\begin{document}
\begin{titlepage}
\begin{flushright}
{\small 
IFUM 887-FT \\
\texttt{hep-th/0701246}}
\end{flushright}
\vspace{1.0cm}
\begin{center} 
\noindent 
\Large 
{\bf Adding D7-branes to the Polchinski-Strassler gravity background}
\end {center}
\large
\vspace{1ex}
\renewcommand{\thefootnote}{\fnsymbol{footnote}}

\vspace{1cm}
\centerline{Riccardo Apreda ${}^a$, Johanna
Erdmenger ${}^{a,b}$, Dieter L\"ust ${}^{a,b}$, Christoph Sieg ${}^c$
\footnote[1]{\noindent \tt apreda@mppmu.mpg.de, \\
\hspace*{6.3mm}jke@mppmu.mpg.de, \\
\hspace*{6.3mm}luest@mppmu.mpg.de, \\ 
\hspace*{6.3mm}csieg@mi.infn.it .\\}}

\vspace{1cm}
\begin{center}
\emph{ $^a$ Max Planck-Institut f\"ur Physik (Werner-Heisenberg-Institut),\\ 
 F\"ohringer Ring 6, 80805 M\"unchen, Germany}
\\
\vspace{0.2cm}
\emph{$^b$ Arnold-Sommerfeld-Center for Theoretical Physics, Department f\"ur Physik,
Ludwig-Maximilians-Universit\"at, Theresienstra\ss e 37, 80333 M\"unchen, Germany}\\
\vspace{0.2cm}
\emph{$^c$  Dipartimento di Fisica, Universit\`a degli Studi di Milano, \\
Via Celoria 16, 20133 Milano, Italy}
\end{center}

\medskip
 
\rm
\abstract
\noindent
We motivate and summarize our analysis of hep-th/0610276
in which we consider $\text{D}7$-brane probe embeddings
in the Polchinski-Strassler background with $\mathcal{N}=2$ supersymmetry. 
The corresponding dual gauge theory is given by the $\mathcal{N}=2^*$ 
theory with fundamental matter. Based on a talk 
given by C.\ Sieg at the RTN Workshop 2006, Napoli.    
\normalsize 

\vfill
\renewcommand{\thefootnote}{\arabic{footnote}}
\end{titlepage} 

\section{Introduction}

The $\AdS/\text{CFT}$ correspondence \cite{Maldacena:1997re} conjectures 
a duality between a string theory in a curved background with a boundary and 
a gauge theory that lives on this boundary.
In a particular case, 
type $\twob$ string theory in $\AdS_5\times\text{S}^5$ with $N$ units of 
Ramond-Ramond $5$-form flux 
is conjectured to
 be dual to the $\mathcal{N}=4$ supersymmetric Yang-Mills (SYM) theory 
with gauge group $SU(N)$, living on the four-dimensional boundary of the 
string background.  
In terms of the 't Hooft coupling constant
$\lambda=g_\text{YM}^2N$, the gauge theory is strongly 
coupled when the string theory reduces to supergravity. The latter is  
seen from $\lambda=\frac{R^4}{\alpha'^2}$ which relates $\lambda$ 
to the curvature radius $R$ of 
$\AdS_5\times\text{S}^5$ and the squared string length $\alpha'$. 
To obtain classical supergravity requires also that the 
$N\to\infty$ limit is taken with $\lambda$ held fixed. This sends the 
string coupling constant $g_\text{s}$ to zero, as 
follows with $g_\text{YM}^2=4\pi g_\text{s}$ from the definition of $\lambda$.
The gauge theory becomes planar in this limit. 

The $\mathcal{N}=4$ SYM theory is not confining but superconformal, and 
moreover it only contains fields that transform in the adjoint representation 
of the gauge group. 
To make contact with the gauge theories that enter the standard 
model of particle physics, and in particular with the confining QCD, one 
therefore has to go beyond the above described setup. 
To find a theory that resembles QCD, we should break the conformal symmetry, 
(some of the) supersymmetries, and add (quark) fields that transform in the 
fundamental representation of the gauge group. The latter can be realized 
by adding D-brane probes \cite{Karch:2002sh}.
The assumption that also in cases with less symmetries a 
$\text{gauge}/\text{gravity}$ correspondence holds 
goes beyond the original conjecture. However, believing in its validity, 
we can continue using it to construct gravity duals to more realistic gauge
theories.

A particularly interesting example for a deformed supergravity 
background which displays a rich physical structure at low energies is
due to Polchinski and Strassler \cite{Polchinski:2000uf}. As far as
the $\text{D}7$-brane probe embedding for the addition of fundamental matter 
is concerned, this background has the appealing feature that
in the far UV, its dual field theory returns again to four-dimensional
$\N=4$ SYM. This allows for the identification of boundary field theory
operators from the asymptotic behaviour of the embedding functions, in
analogy to the standard AdS/CFT dictionary. 

\section{The Polchinski-Strassler background}

The Polchinski-Strassler (PS) background \cite{Polchinski:2000uf} is obtained 
as the gravity dual to $\mathcal{N}=4$ SYM theory, deformed by 
adding mass terms 
$\frac{1}{g_\text{YM}^2}m_p\tr\Phi_p^2$, $p=1,\dots,3$ for the three adjoint
chiral $\mathcal{N}=1$ supermultiplets $\Phi_p$, 
keeping massless only the $\mathcal{N}=1$ gauge supermultiplet. 
The dimensionful
mass parameters $m_p$ clearly break conformal invariance. How much of the 
supersymmetry is broken depends on the concrete values of the $m_p$.
Polchinski and Strassler \cite{Polchinski:2000uf} have discussed the case
$m_1=m_2=m_3$. The corresponding gauge 
theory, in which one of the four supersymmetries is preserved, is known as the 
$\mathcal{N}=1^*$ theory. 
Instead, if mass terms are added for only two of the three chiral multiplets,
e.g.\ $m_1=0$, $m_2=m_3$, the theory preserves two supersymmetries and is 
called $\mathcal{N}=2^*$ theory.

On the gravity side the modifications correspond to the presence of
$3$-from flux $G_3$ built with an imaginary anti-selfdual (IASD) 
$3$-tensor $T_3$, i.e.\ $(\star_6+i)T_3=0$. The effect of $G_3$ on 
the gravity background is better understood if one considers the 
background generating stack of $N$ $\text{D}3$-branes before the 
decoupling limit for the $\text{gauge}/\text{gravity}$ correspondence is
taken. The backreaction of $G_3$ on the $\text{D}3$-branes is an  
example of the dielectric effect found by Myers \cite{Myers:1999ps}, 
i.e.\ it can be interpreted as a polarization of the $\text{D}3$-branes into 
their transverse directions to extended brane sources. 
The near horizon region, into which the decoupling limit zooms,
is no longer given by
$\AdS_5\times\text{S}^5$ -- in fact, its full form is not known
explicitly. However, it is asymptotically 
$\AdS_5\times\text{S}^5$. Therefore, at sufficiently large distance from 
the sources one can treat the extension of the sources perturbatively, such 
that the background can be computed as a perturbation series around  
$\AdS_5\times\text{S}^5$. 

The background fields receive contributions from the
backreaction of the non-vanishing $3$-form flux $G_3$ on the gravity 
background. The linear or oder $\mathcal{O}(m)$ contributions are
discussed in the original paper \cite{Polchinski:2000uf}, where it is
shown that apart from the presence 
of the $2$-form potentials $\tilde C_2$ and $B$ of $G_3$, also $6$-form 
potentials are induced. At quadratic order $\mathcal{O}(m^2)$ the dilaton 
\cite{Polchinski:2000uf,Freedman:2000xb} as well as the metric and 
$4$-form potential $C_4$ are modified \cite{Freedman:2000xb}.
The quadratic order background has been completed in \cite{Apreda:2006bu} 
with the $8$-form potential $C_8$.
In terms of the potentials $\tilde C_2$ and $B$ of $G_3$ we have found
\begin{equation}
\begin{aligned}\label{bgforms}
C_8&=-\frac{1}{6}\big(
\e^{2\hat\phi}\tilde C_2\wedge\tilde C_2+B\wedge B\big)\wedge\hat C_4 \col
\end{aligned} 
\end{equation}
where $\hat\phi$ and $\hat C_4$ denote the uncorrected dilaton 
and $4$-form potential, respectively. 
At cubic order $\mathcal{O}(m^3)$ it has been shown \cite{LopesCardoso:2004ni}
that the flux $G_3$ itself is corrected in such a way that in the dual 
gauge theory a gaugino condensate is generated.

We have analyzed the symmetry structure of the $\mathcal{N}=2$ PS
  background and found that the $\text{D}3$-branes in the deep interior
  are polarized such that they form two overlapping spheres,
\begin{equation}
(y^5)^2 +  (y^6)^2 +  (y^7)^2 = r_0{}^2\, , \qquad
   (y^7)^2 +  (y^8)^2 +  (y^9)^2  = r_0{}^2\, ,
\end{equation}
where the $y^i$ are the six directions perpendicular to the $\text{D}3$-branes
(see also Table 1), and $r_0$ is a constant that depends on the polarizing potential $B$. This geometry preserves an $SU(2)\times SU(2)
\simeq SO(4)$ symmetry very suitable for the embedding of a $\text{D}7$-brane.
We see that there is no polarization in the $y^4$ direction.
Let us remark that close to the extended $\text{D}3$-brane 
sources with extension $r_0$ the perturbation theory breaks down, and the 
background is not known.

\section{$\text{D}7$-brane probes}

The original $\mathcal{N}=4$ SYM theory and hence also its mass deformations 
only contain fields that transform in the adjoint representation of the gauge 
group. Karch and Katz \cite{Karch:2002sh} have proposed to add 
$N_\text{f}$ field flavours which transform in the
fundamental representation (henceforth denoted as quarks) 
by embedding $N_\text{f}$ spacetime-filling $\text{D}7$-branes
into $\AdS_5\times\text{S}^5$. In the brane picture, the $N_\text{f}$ quark 
flavours correspond to open strings with one of their endpoints sitting on 
the stack of $\text{D}3$-brane and the other one ending on one of the 
$N_\text{f}$ $\text{D}7$-branes. The choice $N_\text{f}\ll N$ thereby 
allows one to neglect the backreaction of the $\text{D}7$-branes on the 
background, i.e.\ to consider them as brane probes. 

It turns out that a $\text{D}7$-brane probe
spans an $\AdS_5\times\text{S}^3$ inside $\AdS_5\times\text{S}^5$.
Along the holographic direction $r$ of $\AdS_5\times\text{S}^5$ the brane
fills all of $\AdS_5$ from the boundary at $r\to\infty$ 
down to a minimal value $r=\hat u$ at which it terminates. 
In the $\AdS/\text{CFT}$ correspondence the radial coordinate $r$ has 
the interpretation of an energy scale with small and large $r$ 
corresponding to the IR and UV regimes in the dual gauge theory.
Therefore, the termination of the $\text{D}7$-brane 
at $r<\hat u$ means that there exists an energy threshold below which 
the corresponding quark degree of freedom cannot be excited. 
The threshold is proportional to the quark 
mass $m_\text{q}$. The precise relation reads 
\begin{equation}\label{mqinhatu}
m_\text{q}=\frac{R}{2\pi\alpha'}\hat u\pnt
\end{equation}

Adding a $\text{D}7$-brane probe with $m_\text{q}=0$ clearly 
breaks the $SO(6)$ isometries of the $\text{S}^5$ to $SO(4)\times U(1)$.
In the dual gauge theory this corresponds to a breaking of the 
$SU(4)$ R-symmetry to $SU(2)\times U(1)$. The corresponding gauge theory is
$\mathcal{N}=2$ SYM with massless fundamental matter, which is still a 
superconformal theory. 
In addition, for $m_\text{q}\neq0$ the $U(1)$ factor, 
which rotates the directions transverse to the $\text{D}7$-brane probe, 
is broken. The dual gauge theory is $\mathcal{N}=2$ SYM with fundamental matter
with mass $m_\text{q}$. This theory has an $SU(2)$ R-symmetry.
The fundamental mass breaks the conformal symmetry.


In \cite{Apreda:2006bu} we have studied the embedding of $\text{D}7$-brane 
probes into the $\mathcal{N}=2$ Polchinski Strassler background. 
We have chosen the 
$\mathcal{N}=2$ case with masses $m_1=0$, $m_2=m_3$ 
instead of the $\mathcal{N}=1$ case, since this background 
preserves an $SO(4)$ symmetry in the transverse directions.
An embedding of the $\text{D}7$-brane probe as shown in table 
\ref{tab:D3D7embed}
\begin{table}[t]
\begin{center}
\begin{tabular}{|c|c|c|c|c|c|c|c|}
\hline
& $x^\mu$ & $y^4$ & $y^5$ & $y^6$ & $y^7$ & $y^8$ & $y^9$ \\
\hline
$\text{D}3$ & $+$ & $-$ & $-$ & $-$ & $-$ & $-$ & $-$ \\
$\text{D}7$ & $+$ & $-$ & $+$ & $+$ & $-$ & $+$ & $+$ \\
\hline
\end{tabular} 
\caption{Orientation of the background generating $\text{D}3$ branes 
and of the $\text{D}7$-brane probe in the $10$-dimensional spacetime.}
\label{tab:D3D7embed}
\end{center}
\end{table}
then allows the $\text{D}7$-brane embedding coordinates to 
depend on the radial variable $\rho$ 
in the four parallel directions $y^5$, $y^6$, $y^8$, $y^9$ only. 
These directions are dual to the complex
scalars of the massive chiral multiplets $\Phi_2$, $\Phi_3$ of the gauge 
theory. The two embedding directions $y^4$ and $y^7$ correspond to the 
complex scalar of the massless $\Phi_1$.

In order to be able to find analytic solutions for the embeddings,
we have decomposed the embedding coordinates $y^4$, $y^7$ themselves 
into the constant unperturbed embedding coordinates $\hat y^m$ in 
$\AdS_5\times\text{S}^5$, and a non-constant correction $\tilde y^m$ 
according to 
\begin{equation}\label{embedexp}
y^m(\rho)=\hat y^m+\tilde y^m(\rho)\col\qquad m=4,7\pnt
\end{equation}
The boundary conditions are fixed by
$y^m(\rho\to\infty)=\hat y^m$ and by the requirement that the corresponding 
solution is regular at $\rho=0$. The unique analytic solution is valid 
whenever $\hat y^m\gtrsim\tilde y^m$ which according to 
\eqref{mqinhatu} requires a sufficiently 
large quark mass $m_\text{q}$.
 
The constant $\text{D}7$-brane probe embeddings $\hat y^m$ in pure 
$\AdS_5\times\text{S}^5$ 
acquire corrections $\tilde y^m$ not before the quadratic order 
$\mathcal{O}(m^2)$ in the $G_3$-form flux perturbation, such that we were 
forced to work at least at this order. We therefore had to 
deal with a complicated backreaction that e.g.\ 
affects the metric such that it becomes non-diagonal \cite{Freedman:2000xb}.
We also completed the known corrections
to the dilaton, metric and $4$-form with the $8$-form $C_8$ given in 
\eqref{bgforms}. It also 
is of order $\mathcal{O}(m^2)$, and it couples to the worldvolume of the 
$\text{D}7$-brane probes. 

For the $\text{D}7$-brane probe oriented as shown in table \ref{tab:D3D7embed}
we have then computed the action as the sum of the Dirac-Born-Infeld and 
Chern-Simons action. From this result we extracted the equations of motion 
for the corrections $\tilde y^m$.

\begin{figure}[t]
\begin{center}
\epsfig{file=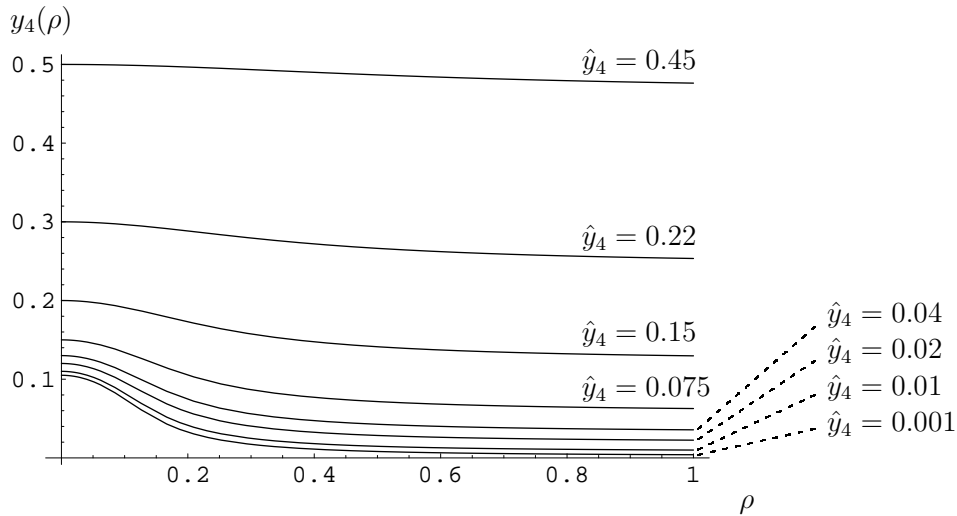, width=0.8\textwidth}
\vspace{3mm}
\caption{Embeddings along $y_4$ with different boundary values $\hat y^4$.
The background generating $\text{D}3$-branes are not polarized in this 
direction. 
The AdS radius has been set to $R=1$ and the adjoint deformation to
$m=0.2$. Lengths are dimensionless and measured in units of $R$.}
\label{y4embeddings}
\end{center}
\end{figure}
We have found that the embedding with $y^4=y^7=0$ is not corrected up to 
order $\mathcal{O}(m^2)$. Even if this embedding enters the regime where the 
perturbative expansion of the background breaks down, we expect that it is 
preserved also in the full solution. 
Furthermore, the two embeddings with either $y^7=0$ or $y^4=0$ are singled 
out: the coordinate that is zero is not corrected 
up to order $\mathcal{O}(m^2)$.
These embeddings with $y^m=(u,0)$ and $y^m=(0,u)$ are perpendicular to or 
respectively contain this direction into which the original background 
generating $\text{D}3$-branes are polarized.

Applying the method of the holographic renormalization 
\cite{Bianchi:2001kw,Skenderis:2002wp,Karch:2005ms}
we have found that the regularized on-shell action can be made to vanish by 
choosing an appropriate scheme, i.e.\ by adding appropriately 
chosen finite counterterms. In particular, this demonstrates the 
absence of a quark condensate, which in a generalization of the
  standard $\AdS/\text{CFT}$ dictionary would follow from the $\rho$ dependence
of the embedding. The absence of the quark condensate
is a necessary condition for 
supersymmetry to be preserved. 

We have also studied the two different embeddings in $y^4 $ and $y^7$ 
direction numerically.
Our findings for the embeddings along $y^4$ with different boundary values
$\hat y^4$ are shown in figure 
\ref{y4embeddings}.
We see that the embeddings are repelled from the singularity in 
this direction, and that for larger boundary values $\hat y^4$ and therefore
according to \eqref{mqinhatu} for
larger quark masses $m_\text{q}$ the embeddings approach the constant 
embedding in $\AdS_5\times\text{S}^5$. 
On the other hand, the embeddings in the $y^7$ direction feel the
effect of the brane shell. Although our ${\cal O}(m^2)$ approximation
breaks down at scales of the order of this expected brane shell, we see
evidence in our numerical result that the D7 brane merges with this
shell and thus is repelled much stronger from the origin
than the brane embedded in the $y^4$ direction. We refer the reader to 
\cite{Apreda:2006bu}  for details about the $y^7$ embedding.

To obtain the meson mass $M$ as a function of the quark mass $m_\text{q}$ 
\cite{Kruczenski:2003be} we have numerically analyzed the fluctuations 
around the embedding along $y^7$.
In the regime where the expansion of the gravity background 
is valid it is of the form $M^2=b\,m_\text{q}^2+c$, where $b$ and the mass gap 
$c$ are some constants. This supergravity result agrees with field
  theory expectations: The presence of the adjoint masses $m$ 
leads to a mass gap in the meson spectrum $M$.
Moreover, the above functional form does not contain a dependence on the 
quark bilinear 
since this would appear as a linear dependence on $m_\text{q}$. 
Therefore, the meson spectrum
found also supports our holographic renormalization result
that the quark condensate vanishes and hence supersymmetry 
can be preserved.

\section*{Acknowledgements}
\addcontentsline{toc}{section}{Acknowledgements}
The work of C.\ S.\ is supported by INFN as `assegno di ricerca'.
 R.\ A.\ has been supported by DFG, grant ER301/2-1.
\small
\bibliographystyle{utphys}
\addcontentsline{toc}{section}{Bibliography}
\bibliography{references}

\end{document}